\documentclass[prl,twocolumn,amsfonts,showpacs]{revtex4-1}

\pdfoutput=1

\usepackage{epsfig}
\usepackage{psfrag}
\usepackage{amsmath}
\usepackage{amssymb}
\usepackage{color}
\usepackage{hyperref}
\begin{document}

\author{Chuan-Yin Xia$^{1,2}$}
\author{Hua-Bi Zeng$^{1}$}
\email{hbzeng@yzu.edu.cn}
\author{Hai-Qing Zhang $^{3}$}
\email{hqzhang@buaa.edu.cn}
\author{Zhang-Yu Nie$^2$}
\author{Yu Tian$^{4,1}$}
\author{Xin Li$^4$}

\affiliation{$^1$ Center for Gravitation and Cosmology, College of Physical Science
and Technology, Yangzhou University, Yangzhou 225009, China}
\affiliation{$^2$ School of Science, Kunming University of Science and Technology, Kunming 650500, China}
\affiliation{$^3$ Center for Gravitational Physics, Department of Space Science \&
 International Research Institute for Multidisciplinary Science, Beihang University, Beijing 100191, China}
\affiliation{$^4$ School of Physical Sciences, University of Chinese Academy of Sciences, Beijing 100049, China  \& Institute of Theoretical Physics, Chinese Academy of Sciences, Beijing 100190, China}

\title{Vortex Lattice in a Rotating Holographic Superfluid}

\begin{abstract}
By utilizing the AdS/CFT correspondence, we explore the dynamics of strongly coupled superfluid vortices in a disk with constant angular velocity at a finite temperature. Each vortex in the vortex lattice is quantized with vorticity one from the direct inspection of their phases. As the angular velocity of the disk is greater than a critical value, the first vortex will be excited as expected from theoretical predictions. The subsequent two and more vortices are also generated by increasingly rotating the disk, resulting in the remarkable step transitions for the angular velocity to excite each individual vortex. When the vortex number is large enough, the density of vortices is found to be linearly proportional to the angular velocity, which matches the Feynman relation very well.  We also find that varying the temperature does not alter this Feynman relation.
\end{abstract}
%\pacs{ 05.30.-d;73.43.-f;05.30.Pr}
\maketitle
\pagebreak
%\section{Introduction}
%The description of the out of equilibrium dynamics of strongly
%interacting quantum systems remains one of the most challenging
%problems in theoretical physics.

\emph{Introduction.}---
Quantized vortex has a profound effect on the behavior of Type-II superconductors and superfluids.
The quantized circulation is a macroscopic quantum mechanical effect, which is a direct
consequence of a single-valued  wave function, where the phase must change by
$2\pi n$ ($n\in Z$ is the vorticity) around a vortex core.  Theoretical studies in equilibrium states predicted that the vortex lines would form a stable triangle lattice  minimizing the
free energy in both superconductor \cite{Abrikosov} and superfluid \cite{Tkachenko}.
In experiments, vortex lines has been observed in both helium II in rotating containers \cite{Packard,Yarmchuk,Bewley} as well as Bose-Einstein condensation (BEC) in cold atoms \cite{Madison,Shaeer}.
The vortex lattice formation and vortex phase diagram under a constant  rotation of a container can be simulated by solving the powerful Gross-Pitaevskii (GP) equation numerically \cite{Pethick,Fetter,Kasamatsu1,Kasamatsu2}.
However, GP equation is only valid for a weakly coupled system at zero temperature, studies on the vortex formation in a strongly coupled  superfluid at finite temperature is still lacking.

 The AdS/CFT (Anti-de Sitter/Conformal Field Theory) correspondence \cite{Maldacena,Gubser,Witten}
provides a complete  description  ``valid  at  all  scales''  of  a  strongly  interacting quantum many-body system in terms of a classical gravitational system at finite temperature \cite{Liu}.
 The holographic study of  superfluid (or superconductor) was originally introduced in \cite{Gubser2008,Hartnoll,Herzog},
where the $U(1)$ gauge symmetry is spontaneously broken in a AdS planar black hole background.
Later, the one vortex solution of this holographic model was obtained in both superfluid and superconductor in \cite{Montull,Keranen,Dias,Wu}. Triangle lattice solution of vortex lattice were found from perturbative calculations near critical point in \cite{Maeda}.  Holographic superconductor/superfluid model also gives insights to the dynamics of a continuous
phase transition even in far from equilibrium dynamics \cite{Murata,Bhaseen,Li,Bai,Tianyu}, the nonlinear response to
a strong external field \cite{Zeng:2016api,Zeng:2016gqj}, as well as the critical behavior of a non-equilibrium phase transition \cite{Zeng:2018ero}. %Extending the equations to the cases with all coordinates dependent gives
%a systematic new and first-principle approach to study quantum turbulence \cite{Liu,Du:2014lwa} as well as to the topological defects formation predicted by the Kibble-Zurek mechanism\cite{Chesler:2014gya,Sonner:2014tca}.

In this letter, we investigate the formation of the vortex lattices in a strongly coupled holographic superfluid in a constantly rotating disk with angular velocity $\Omega$ when the temperature is away from zero. The quantized vortices were found by direct inspection of the  phases of scalar field. We also get the step transitions of the critical angular velocities $\Omega_c$'s which will excite each individual vortex as the vortex number is relatively small. As the vortex number is large enough, the density of the vortices is found to be linearly proportional to the angular velocity, which is consistent with Feynman relation \cite{Feynman}.

\emph{Holographic Model}.---
A simple action for holographic superfluid can consist of a complex scalar field $\Psi$ with mass $m$, minimally coupled to a $U(1)$ gauge field $A_\mu$ \cite{Gubser2008,Hartnoll,Herzog}, %(We have set $c=\hbar=1$.)
\begin{equation}
S=\int  d^4x \sqrt{-g}\Big[-\frac{1}{4}F^2-|D\Psi|^2-m^2|\Psi|^2\Big],
\label{model}
\end{equation}
where $F_{\mu\nu}=\partial_\mu A_\nu-\partial_\nu A_\mu$, $D_\mu=\partial_\mu-iqA_\mu$ with $q$ the charge. The theory can be defined in a $AdS_4$ black hole background with Eddington-Finkelstein coordinates,
\begin{equation}
ds^2=\frac{\ell^2}{z^2}\left(-f(z)dt^2-2dtdz + dr^2+ r^2d\theta^2\right).
\label{metric}
\end{equation}
in which $\ell$ is the AdS radius, $z$ is the AdS radial coordinate of the bulk
and $f(z)=1-(z/z_h)^3$.  Thus, $z=0$ is the AdS boundary while $z=z_h$ is the horizon; $r$ and $\theta$ are respectively the radial and angular coordinates of the dual $2+1$ dimensional boundary, which is a disk in our model. The Hawking temperature is $T=3/(4\pi z_h)$.
For simplicity, the probe limit is adopted in the paper by assuming that the matter fields do not affect the gravitational fields.
The black hole back ground Eq. (\ref{metric}) is static rather than rotating, then the
superfluid is rotating relative to the disk, or equivalently  we can treat the superfluid as a static observer
then the disk is rotating with the fact that  superfluid is of zero viscosity.
Without loss of generality we rescale $\ell = z_h = 1$.
Therefore, the equations of motions (EoMs) can be written as
\begin{equation}
(-D^2+m^2)\Psi=0,~~~  \partial_\mu F^{\mu\nu}=J^\nu,
\label{eom}
\end{equation}
where $J^{\mu}=i (\Psi^*D^{\mu} \Psi-\Psi D^{\mu} \Psi^*)$
is the bulk current.
Proper boundary conditions should be imposed in order to solve the EoMs \eqref{eom}.  For simplicity, the axial gauge $A_z=0$ is adopted as in \cite{Herzog}. We impose the regular boundary conditions of the all physical solutions at the horizon of the black hole.  Explicitly, we set $A_t=0$ at the horizon as in \cite{Natsuume:2017jmu}. Other fields at the horizon can be determined from the previous time steps in the time evolution schemes, rather than imposed by hand.
Near the boundary $z=0$, the general solutions take the asymptotic form as,
\begin{eqnarray}
A_\nu(t,z,r,\theta)&=& a_\nu(t,r,
\theta)+ b_\nu(t,r,\theta) z+\mathcal{O}(z^2),
\label{aboundary} \\
 \Psi(t,z,r,\theta)&=& \Psi_1(t,r,\theta) z+ \Psi_2(t,r,\theta) z^2+\mathcal{O}(z^3).
 \label{psiboundary}
\end{eqnarray}
From the AdS/CFT dictionary, the coefficients $a_{r,\theta}$ can be related to the superfluid velocity along $r, \theta$ directions while $b_{r,\theta}$ as the conjugate currents \cite{Montull}.  Coefficients $a_t$ and $b_t$ are respectively interpreted as chemical potential and charge density in the boundary field theory; $\Psi_1$ is the source term while $\Psi_2$ is the vacuum expectation value $\langle O\rangle$ of the dual scalar operator. In the superfluid phase we always impose $\Psi_1\equiv0$ and $a_t=\mu>\mu_c$ in the $z=0$ boundary. In this letter we choose $m^2=-2$, thus the critical chemical potential is  $\mu_c\sim4.06$.
In order to study the formation of superfluid vortex lattice in a rotating disk, we impose the angular boundary condition in the $z=0$ boundary  as \cite{Domenech}
\begin{equation}
a_\theta=\Omega r^2,
\label{rotation}
\end{equation}
where $\Omega$ is the constant angular velocity of the disk.
While in the rotating  frame of reference,
the field $a_\theta$ represents the relative velocity between the superfluid and the reference of the rotating disk. Therefore, $a_\theta$ is also the velocity of the superfluid seen from the rotating disk.
Then the  most  convenient way to introduce rotation in the holographic superfluid is to assume a static disk on the black hole boundary while the superfluid is rotating relative to the disk. With the fact that the superfluid is incompressible,
the existence of the inertial force can not introduce a superfluid velocity along the radial direction
of the disk, then the physics of a rotating superfluid in a static disk is
   equal to the physics of a static superfluid in a rotating disk.
{One should note that the definition of $\Omega$ above has mass dimension $+2$. Therefore, the physical angular velocity should be scaled by the chemical potential which is the energy scales of this system. By doing this, one can get the correct physical angular velocity with mass dimension $+1$, and avoid the superluminal problems. }  We also impose $a_r=0$ at the boundary since we assume no superfluid flows in the radial direction of the disk. In our model, the radius of the boundary disk is set as $r=R$. { The Neumann boundary conditions are adopted both at $r=R$ and $r=0$, i.e., $\partial_r h_i=0$ where $h_i$ represents all the fields  except $a_\theta$. Please note that this Neumann boundary condition is imposed in the whole range of $z$. }.. The periodic boundary conditions are used along $\theta$ direction, thus, we used the Fourier decomposition in the $\theta$ direction for all the fields. Chebyshev spectral methods are used in the $(z, r)$ direction. Time evolution is simulated by the fourth order Runge-Kutta method.
The initial configuration at $t=0$ is chosen to be a homogenous superfluid state without any rotation at a fixed temperature below $T_c$.

Free energy $F$ of the system can be computed from the renormalized on-shell action $S_{\rm ren.}$, i.e., $F=T S_{\rm ren.}$. The renormalized on-shell action consists of two parts, viz., $S_{\rm ren.}=S_{\rm o.s.}+S_{\rm c.t.}$, in which $S_{\rm o.s.}$ is the bare on-shell action by subtracting the equations of motions from the action \eqref{model} and $S_{\rm c.t.}$ is the counter term to remove the divergence near $z=0$ boundary.  Explicitly, $S_{\rm c.t.}=-\int dtdrd\theta\sqrt{-\gamma}\Psi^*\Psi|_{z=0}$ which is computed near $z=0$ boundary, and $\gamma$ is the determinant of the reduced metric on the boundary surface.
Therefore, the final form of the renormalized on-shell action is, (In the numerical computation we set $q=1$.)
\begin{eqnarray}\label{sonshell}
S_{\rm ren.}&=&-\frac12\int dtdzd\theta  \left[\frac1r A_\theta\partial_rA_\theta\right]\bigg|_{r=R} \nonumber \\
&&\hspace{-1cm}+\frac12\int dt dr d\theta\left[r\left(-a_tb_t+\frac{1}{r^2}a_\theta b_\theta+\Psi^*_1\Psi_2+\Psi^*_2\Psi_1\right)\right]\bigg|_{z=0} \nonumber \\
&&\hspace{-1cm}+\frac{iq}{2}\int dtdzdrd\theta \left[\frac{r}{z^4}A_\mu\left(\Psi^*\partial^\mu\Psi-\Psi\partial^\mu\Psi^*-2iqA^\mu|\Psi|^2\right)\right].~~~~
\end{eqnarray}

%{\red$S_{\rm s.f.}=-\int dtdrd\theta\sqrt{-\gamma}n^zA^tF_{zt}|_{z=0}$} in which $n^z$ is the normal vector of the $z=0$ surface.
%\begin{eqnarray}
%S_{\rm c.t.}&=&-\int dtdrd\theta\sqrt{-\gamma}\psi^*\psi\bigg|_{z=0}\nonumber\\
%&=&-\int dtdrd\theta\left[r\left(\frac{\psi^*_0\psi_0}{z}+\psi^*_0\psi_1+\psi^*_1\psi_0\right)\right]\bigg|%_{z=0},\end{eqnarray}
%\begin{eqnarray}
%S_{\rm s.f.}=-\int dtdrd\theta\sqrt{-\gamma}n^zA^tF_{zt}=\int dtdrd\theta\left(ra_tb_t\right)\bigg|_{z=0}
%\end{eqnarray}
%and define the complex scalar field as {\red $\Psi=|\Psi|e^{in\theta}$} where $n$ is the winding number or vorticity of the superfluid vortex
%-q\int dtdzdrd\theta \left[\frac{|\Psi|^2r}{z^2}\left(\frac{A_\theta}{r^2}(n-qA_\theta)+\frac{q}{f}A^2_t\right)\right].

\emph{Quantized vortex lattice}---
According to Landau's two-fluids model of superfluid \cite{Landau},
the normal components behave like ordinary liquids while the superfluid components
move without dissipation. These two components can have different velocities: $v_n$ for the normal parts and $v_s$ for the
superfluid parts. If the container (a two-dimensional disk in our case)
rotates at a constant angular velocity $\Omega$, the normal component rotates accordingly similar to a rigid
body. This implies the linear velocity $v_n= \Omega \times r$
and the curl $ \nabla \times  v_n= 2\Omega$, in which $r$ is the position vector with
its origin sitting at the vortex core.
In contrast, the superfluid part remains stationary, i.e., $v_s=0 $, at small $\Omega$, which is called the Landau state. However, a stationary liquid in a rotating container implies a higher
free energy. Thus, as $\Omega$ increases to a critical value $\Omega_{c1}$, Landau state becomes
unstable and prefers entering a state with one vortex. Keep increasing $\Omega$ to the second critical
velocity $\Omega_{c2}$, two quantized vortices will appear and locate symmetrically in the disk. Consequently,
higher angular velocities will excite the third, fourth, and subsequent more vortices, which will arrange themselves in the disk according to the minimum of the free energy.

%******************
\begin{figure}[h]
\centering
\includegraphics[trim=5.3cm 10.0cm 4.cm 10.cm, clip=true, scale=0.6, angle=0]{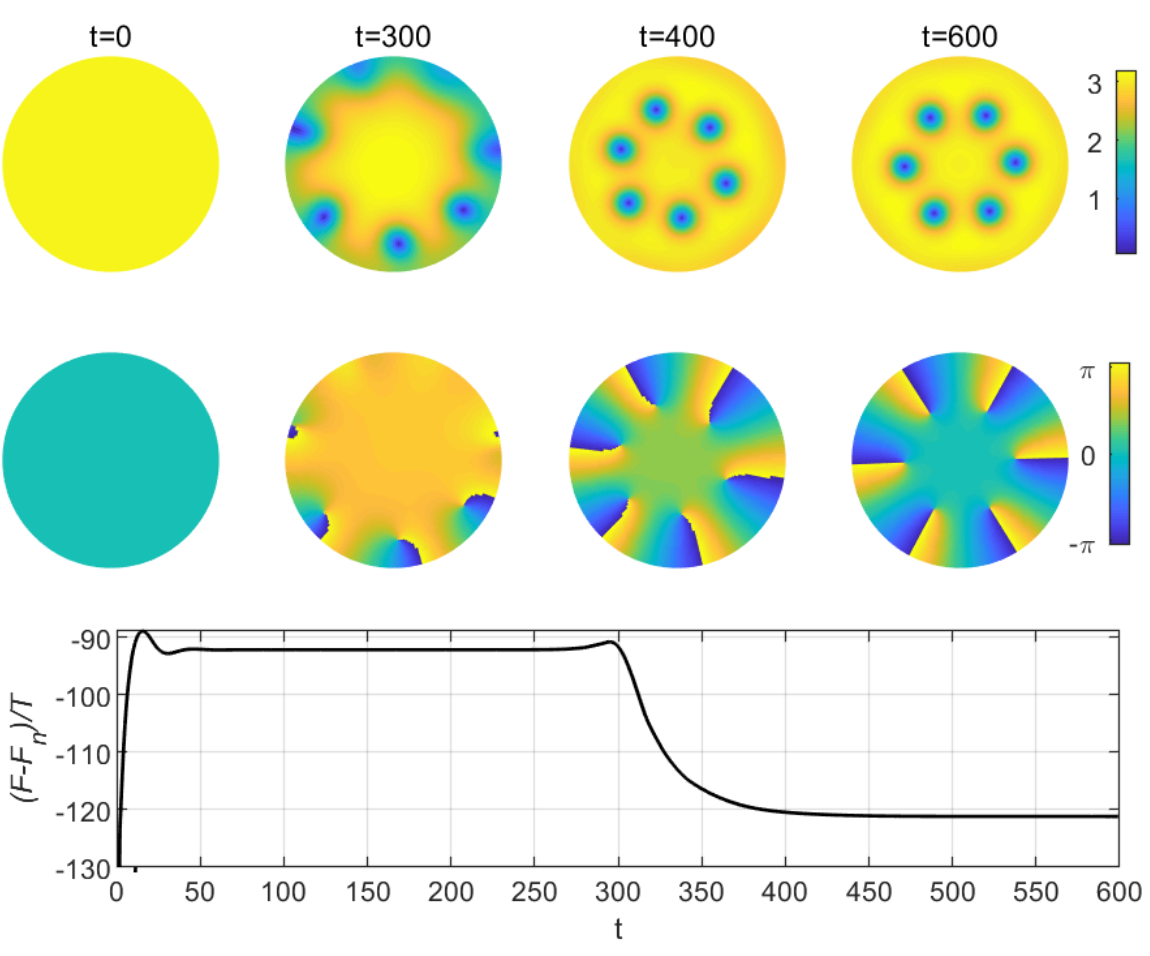}
\caption{\footnotesize{Superfluid vortex lattice formation at separate times (top row), their corresponding phases (middle row) and time evolution of the rescaled free energy (bottom row) for $R=5$ and $\Omega=0.42$. The temperature is chosen to be $T=0.82T_c$.}}\label{fig1}
\end{figure}
%******************

The top row of Fig.\ref{fig1} shows the development of vortex lattices (with $6$ vortices) during time from $t=0$ to final equilibrium state $t=600$ for $R=5$ and $\Omega=0.42$, at the temperature $T=0.82T_c$.
At time $t \sim 300$ the vortices begin to form
from the edge of the disk.  This phenomenon is consistent with the theoretical studies and experiments in literatures \cite{Ruutu}.  At later time vortices will rotate and enter into the inner side of the disk.  The middle row of Fig.\ref{fig1} plots the corresponding phases of the superfluid in top row. In the final stable state, the locations of the vortices can be directly seen from the singularities or branch points of the phases. Circling around the vortex core, the phases vary from $-\pi$ (blue) to $+\pi$ (yellow) with discrepancy $2\pi$, which demonstrates that each vortex is quantized with vorticity $n=1$. The bottom row of Fig.\ref{fig1} shows the time evolution of the corresponding free energy $(F-F_n)/T$, in which $F_n$ is the free energy in the normal state, i.e., $\Psi=0$.  One should note that the free energy formula
is well defined in the equilibrium state, but may not be properly defined in the dynamical case. However, from Fig.\ref{fig1} we see that at least in later time $t>400$ the system is in equilibrium state with a relatively lower free energy, and in the regime $50<t<250$ the system is in a meta stable state with higher free energy. Interestingly, we found a similar profile of time-dependent free energy in \cite{Peeters},
where a superconductor under inhomogeneous magnetic field was studied from condensed matter physics.

%  the free energy will decrease and later saturate into a stable state with six vortices.
% We see that in the beginning of time, free energy will rapidly grow because of rotation of the disk; Then it will stabilize itself for a period of time before the vortices turn out at $t\sim300$. However, this state is metastable with higher free energy, thus, when $t>300$ free energy will decrease to a lower energy. It should be noted that after adding the constant angular velocity to the disk in the beginning, we did not perturb this system any more. Time evolution of the system itself will automatically seek a state with lower free energy. From the bottom row of Fig.\ref{fig1} we see that as $t>300$ the free energy will decrease and later saturate into a stable state with six vortices.

%*****************
\begin{figure}[h]
\centering
\includegraphics[trim=5.3cm 10.2cm 5.4cm 10.3cm, clip=true, scale=0.5, angle=0]{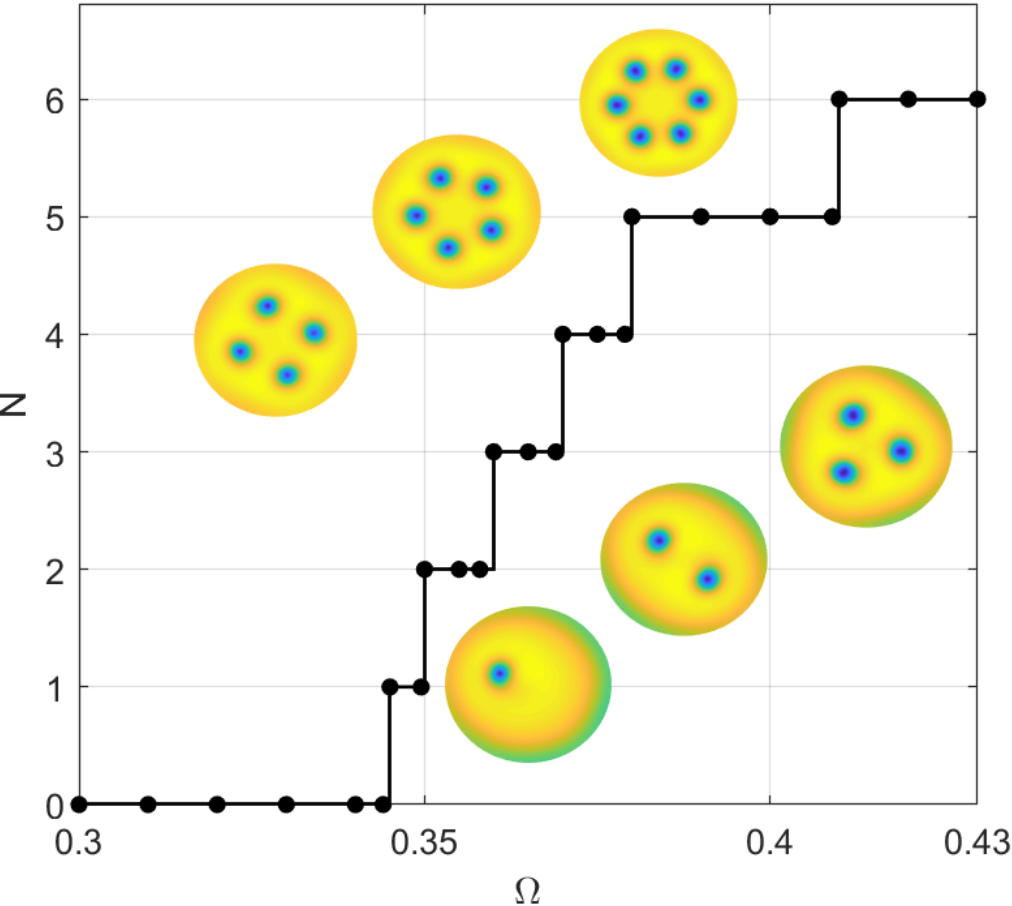}
\caption{\footnotesize{Vortex number $N$ vs. angular velocity $\Omega$ of the disk in the final stable state with $R=5$. The step wise transitions signal the appearance of an additional vortex. The insets are the configurations of the stable vortices from { $N=1$ to $N=6$}. The temperature is $T=0.82T_c$.}}\label{fig2}
\end{figure}
%***************

\emph{Critical angular velocities}--- In Fig.\ref{fig2}, we show the typical step wise relation between the number of vortices $N$ and angular velocity $\Omega$, from $N=1$ to $N=6$ for the case of $R=5$. The corresponding critical angular velocities $\Omega_{c1}, \Omega_{c2} \dots$ can be read from the jump of $N$.
By increasing the angular velocity from zero to $\Omega_{c1}\sim 0.345$, the first vortex will be excited by the rotation of the disk \cite{Landau}. The next two, three and more vortices can also be generated at some larger critical velocities. One possible reason for the unequal spacings is that the appearance of vortex will in some sense break the superfluidity of the superfluid. The normal components of the superfluid will scatter at the vortices, leading to the frictions between the normal and superfluid components inside the disk \cite{Landau}. This friction will also cost some energies which may result in the unequal spacings of the critical angular velocities.  Studies in condensed matter \cite{Hess,Ruutu,Packard} also showed the unequal spacings of the critical angular velocity as vortex number is small.

\emph{Feynman relations}--- For large vortex numbers, the rotations of the superfluid can be regarded as the rotation of a rigid body \cite{Landau}. Therefore, from the path integral of the velocity along the disk circumference (enclosing the whole large number of vortices) and the single-valued phase of the scalar field, one can readily get the Feynman relation \cite{Feynman} as,
\begin{equation}
N= \frac{M\Omega}{\pi } \pi R^2=nA,
\label{linear}
\end{equation}
where  $n\equiv\frac{M\Omega}{\pi} $ is the vortex number density, $M$ is the atom mass of the superfluid and $A\equiv\pi R^2$ is the area of the disk.

%*******************
\begin{figure}[h]
\centering
\includegraphics[trim=3cm 11.2cm 2cm 10.5cm, clip=true, scale=0.6, angle=0]{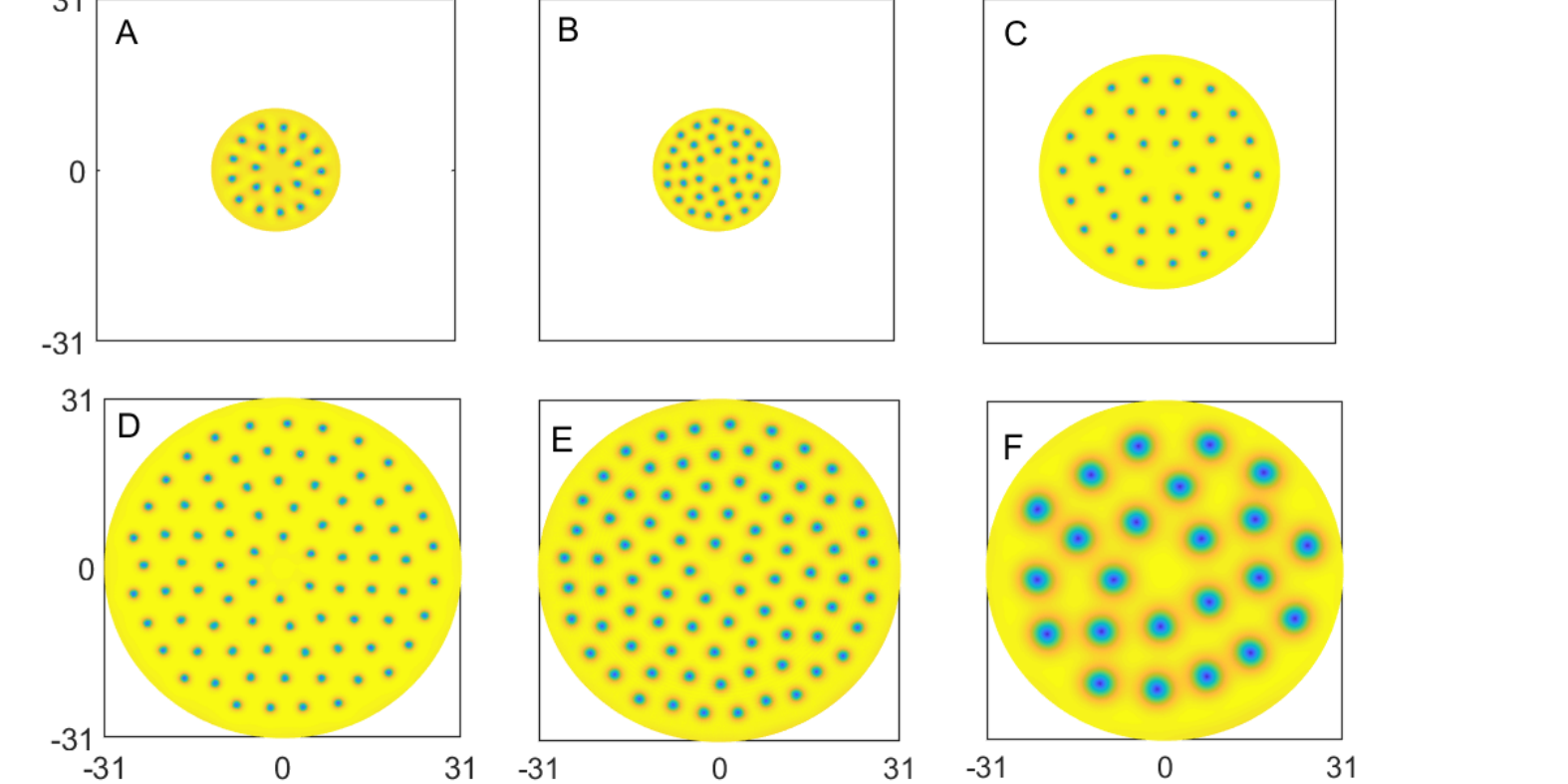}
\caption{\footnotesize{Configurations of a large number of vortex lattices, with vortex number $N=20(A),33(B),37(C), 82(D), 83(E), 23(F)$ corresponding
to $(R, \Omega, T/T_c)=(11, 0.27, 0.82), (11, 0.34, 0.82), $ $(21, 0.1, 0.82), (31, 0.1, 0.82), (31, 0.1, 0.95), (31, 0.03, 0.98)$, respectively.}}\label{fig3}
\end{figure}
%****************

\begin{figure}[h]
\includegraphics[trim=2.8cm 12.6cm 2.8cm 11cm, clip=true, scale=0.55, angle=0]{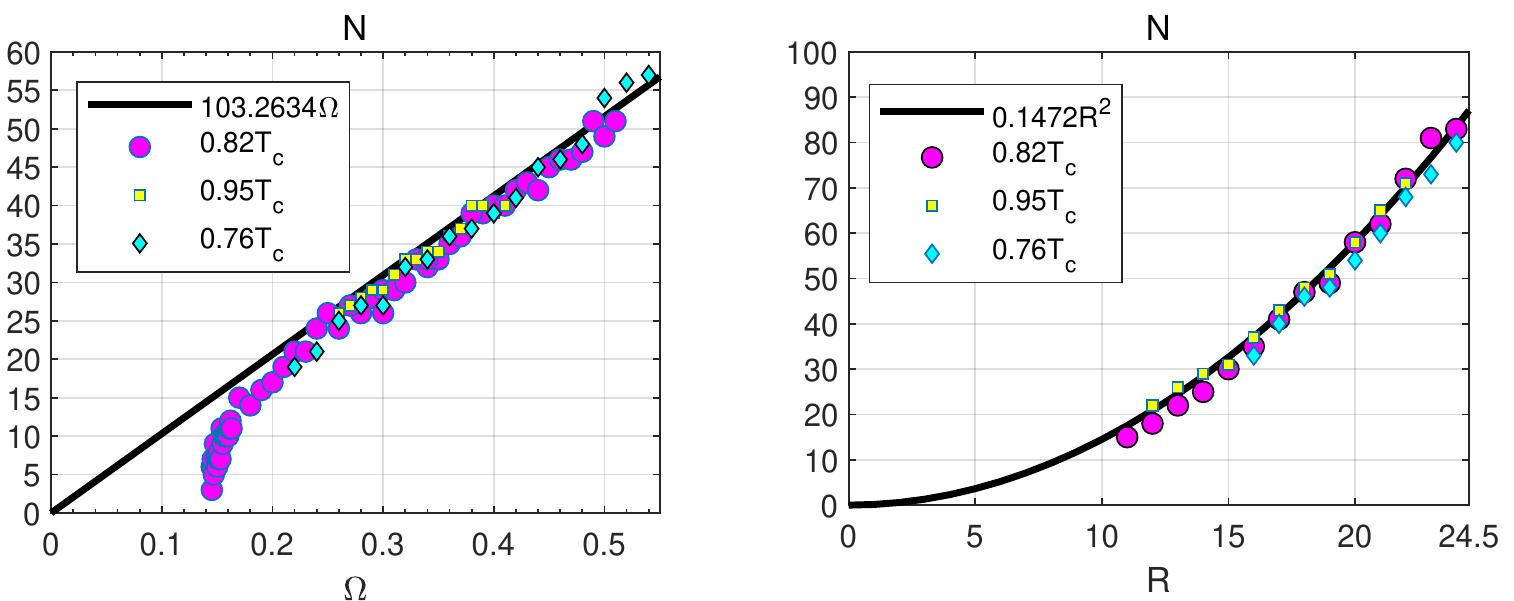}
\caption{\footnotesize{{(Left): Vortex number $N$ vs. angular velocity $\Omega$ under various temperatures with $R=11$. The black line $N\sim103.2634\Omega$ is the linear fitting curve as $N$ is large; (Right): Vortex number $N$ vs. the radius $R$ under various temperatures with $\Omega=0.17$. The black fitting curve is $N\sim0.1472R^2$.}}}\label{fig4}
\end{figure}

In Fig.\ref{fig3} we show the configurations of vortex lattices with large vortex numbers for various $R, \Omega$ and temperatures.
%for $N=20 (A)$, $33 (B)$, $37 (C)$ and $82 (D)$, corresponding to $(R, \Omega)=(11, 0.27)$, $(11, 0.34)$, $(21, 0.1)$ and $(31, 0.1)$ , respectively, at $T=0.82T_c$. And vortex numbers for $N=83 (E)$,corresponding to  $(R, \Omega)=(31, 0.1)$ at $T=0.95T_c$, also $N=23 (F)$  corresponding to  $(R, \Omega)=(31, 0.03)$ at $T=0.98 T_c$. These vortex lattices are uniformly distributed from the symmetry of the disk as well as the minimum of the free energy.
 From the Feynman relation Eq.\eqref{linear}, we may estimate the values of $M$ from the six subfigures in Fig.\ref{fig3}. We obtain $M(A)\sim 0.6122, M(B)\sim0.8021$, $M(C)\sim0.8390$, $M(D)\sim0.8533$, $M(E)\sim0.8637$ and $M(F)\sim0.7978$, respectively. The values of $M$ in subplots $B$, $C$, $D$ and $E$ with more vortices are close since the Eq.\eqref{linear} is valid for large number of vortices. In contrast, in subplot $A$ the vortex number $N=20$ seems not large enough to satisfy the Feynman relation Eq.\eqref{linear}.  For this reason, we plot a large number of vortices with respect to $\Omega$ under various temperatures in the left panel of Fig.\ref{fig4} by fixing $R=11$. One can readily see that as vortex number is relatively large, the relation between $N$ and $\Omega$ under various temperatures are almost in the same line, which indicates that temperature does not alter the Feynman relation.  For large vortex number, the linear relation between $N$ and $\Omega$ is fitted as $N\sim 103.2634\Omega$. Thus, comparing this fitting line with Eq.\eqref{linear}, we can readily get that $M\sim0.8534$. Therefore, the relation Eq.\eqref{linear} becomes
\begin{equation}
N \sim 0.8534 \Omega R^2.
\label{fitting}
\end{equation}
In the right panel of Fig.\ref{fig4}, we show the relation between large number of vortices to the radius $R$ under various temperatures by fixing the angular velocity $\Omega=0.17$. We can also see that different temperatures do not change this relation. The direct fitting of the curve in the  right panel of Fig.\ref{fig4} is $N\sim0.1472R^2$. By comparison, replacing $\Omega=0.17$ into Eq.\eqref{fitting} we get $N\sim 0.1451R^2$, which perfectly matches the fitting (within $1.43\%$ differences) from the right panel of Fig.\ref{fig4}. Therefore, this in turn numerically confirms the Feynman relation Eq.\eqref{linear}!
%****************
%*****************

%******************
\begin{figure}[t]
\centering
\includegraphics[trim=2cm 5.8cm 2.9cm 6.7cm, clip=true, scale=0.22, angle=0]{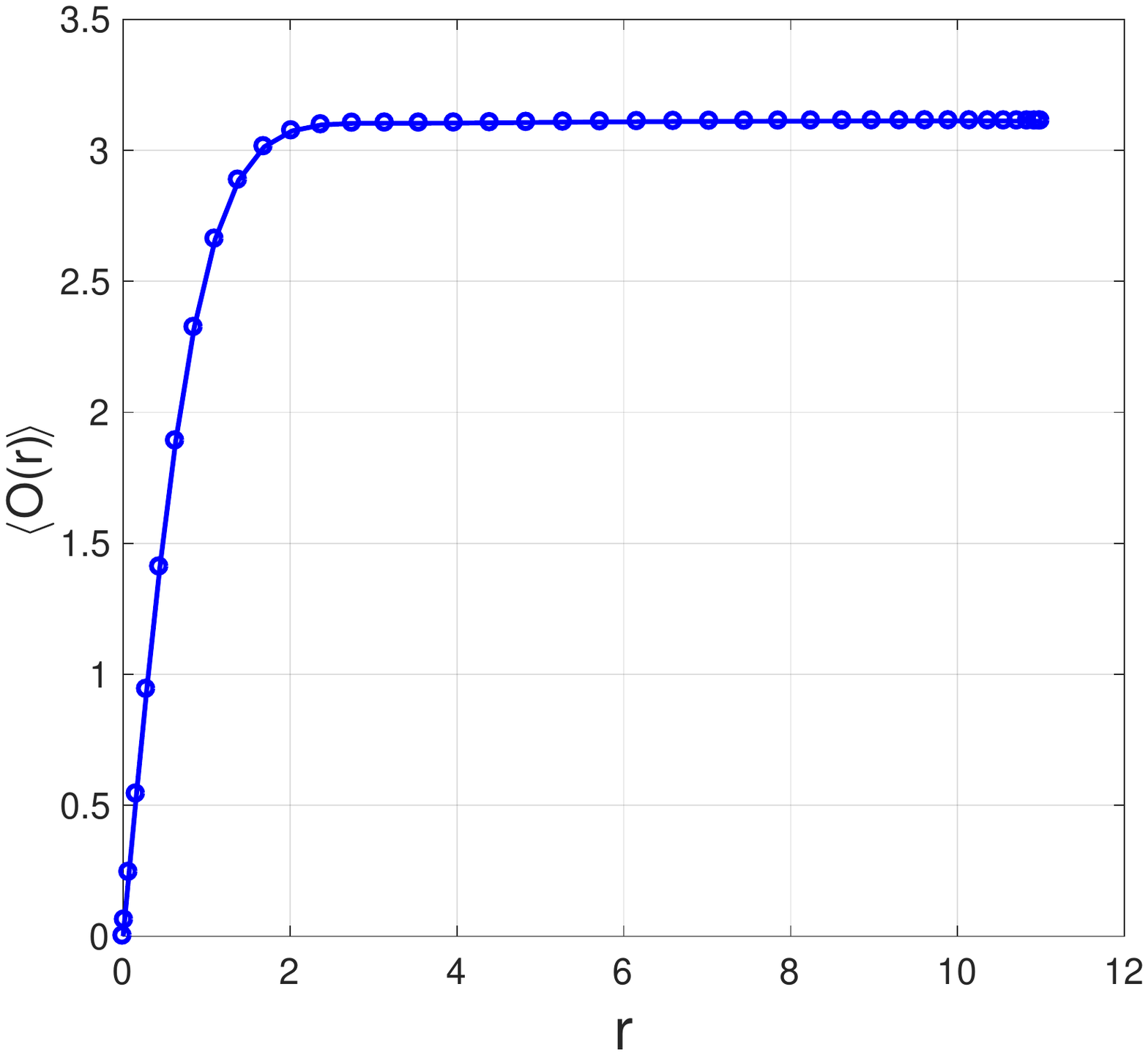}
\includegraphics[trim=1.3cm 6.6cm 2.4cm 7.4cm, clip=true, scale=0.24, angle=0]{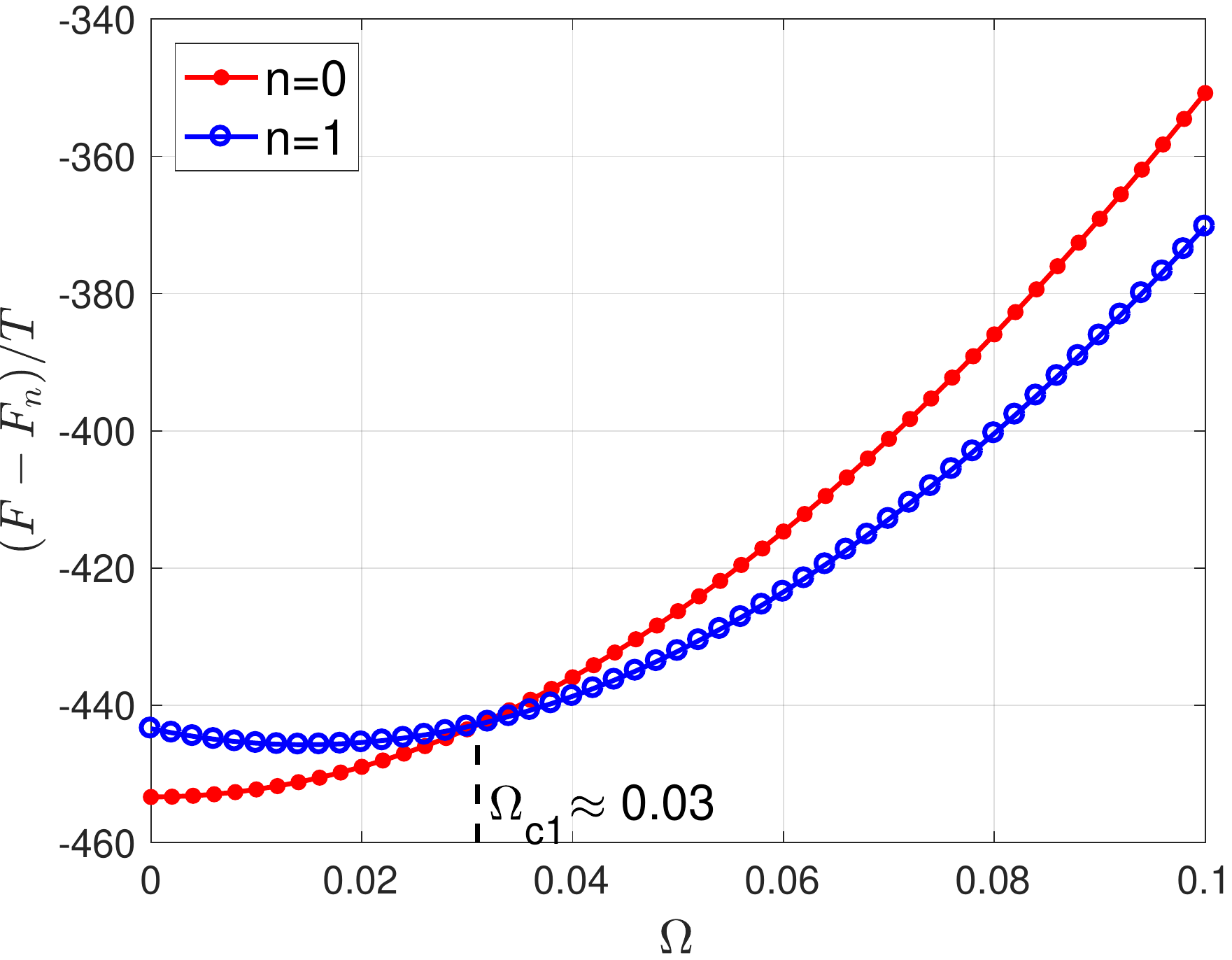}
\includegraphics[trim=2.3cm 6.5cm 2.3cm 7cm, clip=true, scale=0.23, angle=0]{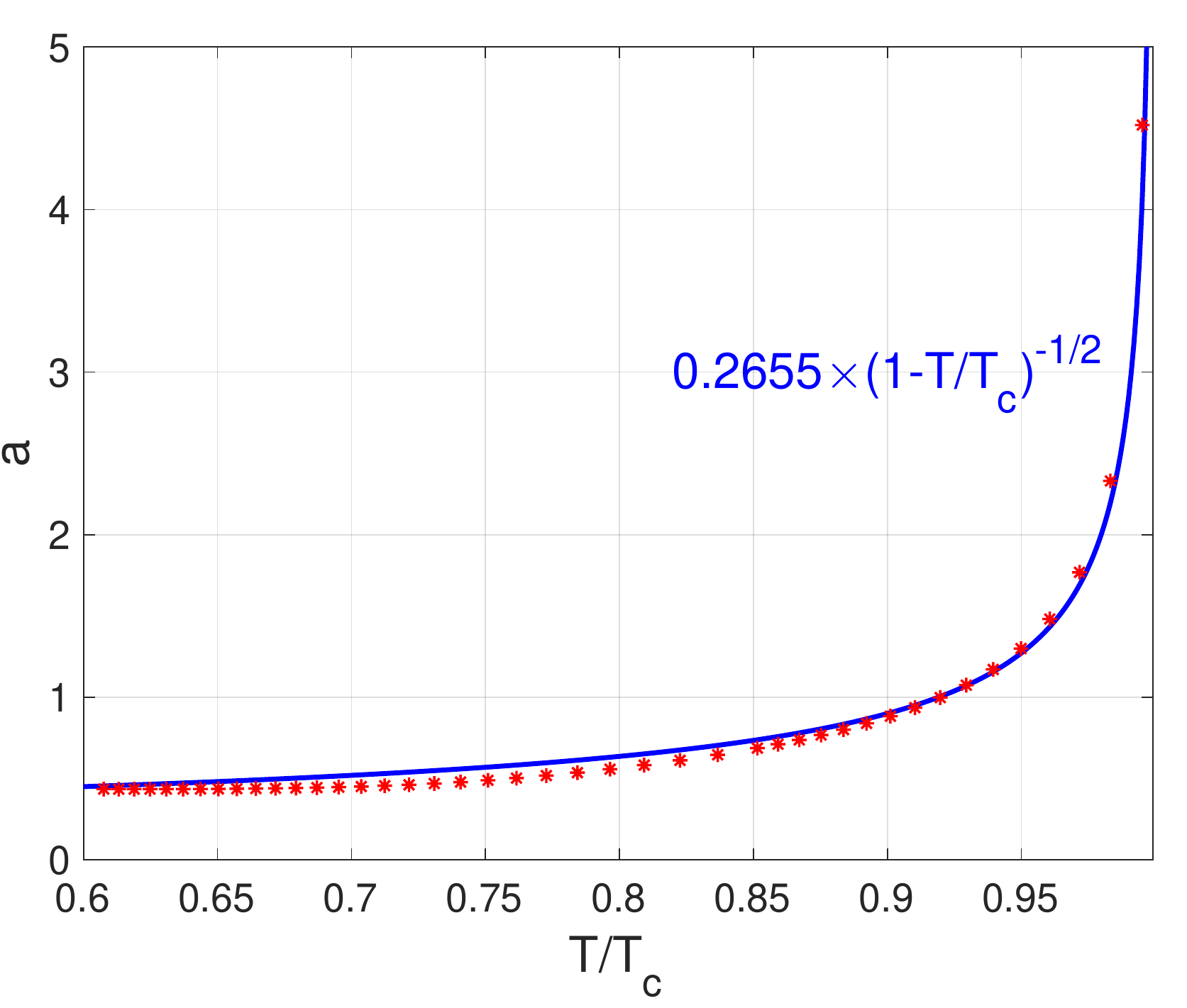}
\includegraphics[trim=1.cm 6.5cm 2.3cm 7cm, clip=true, scale=0.23, angle=0]{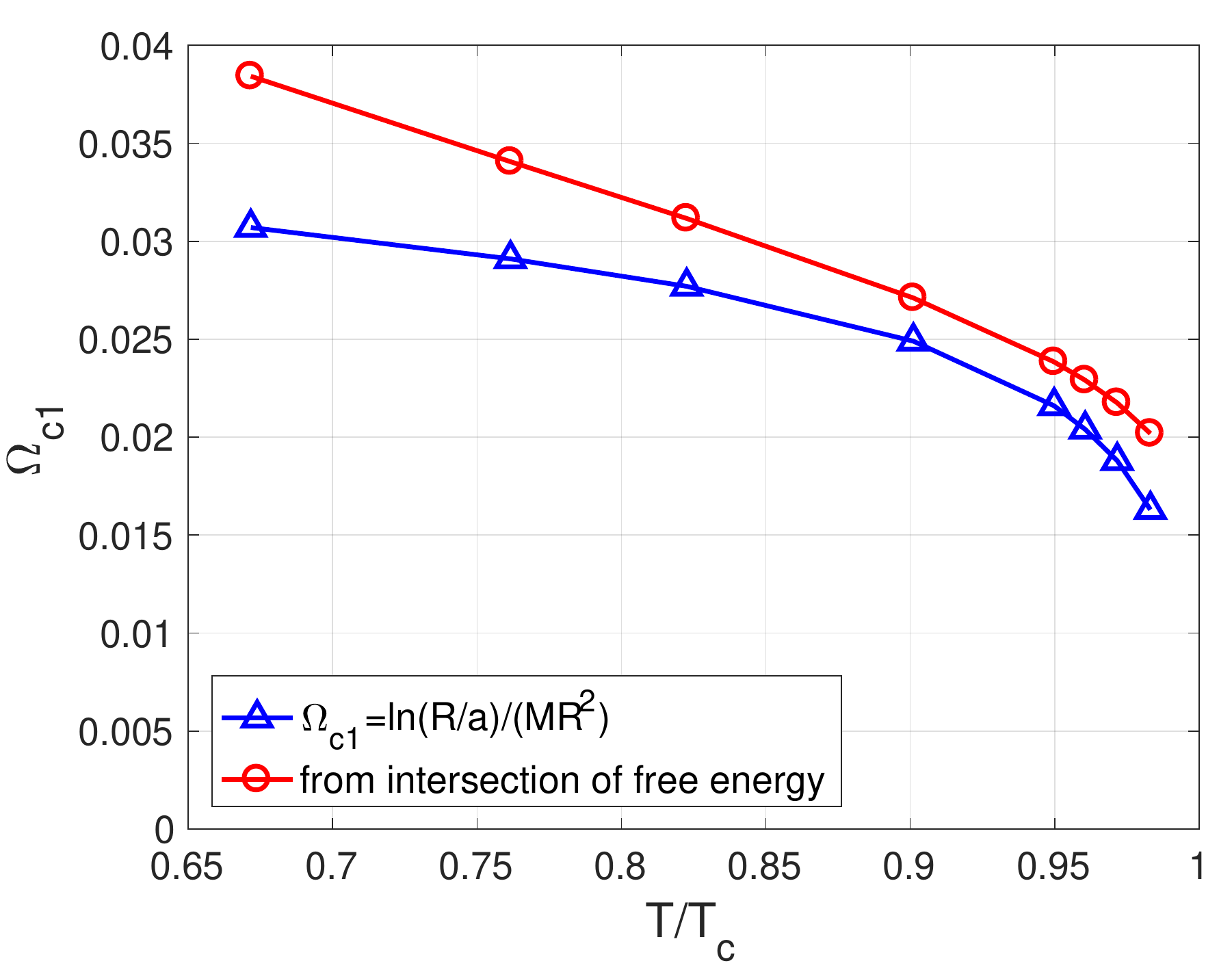}
\caption{\footnotesize{{ (Top Left): Static single vortex solution for $R=11$ at $T=0.82T_c$. The vortex order parameter far from the vortex core is  $\langle O(R)\rangle\sim3.1114$; (Top Right):Free energy vs. the angular velocity for $n=0$ and $n=1$ vortex solutions in static case with $R=11$ at $T=0.82T_c$; (Bottom Left): The vortex size $a$ vs. temperature. The blue line is the fitting curve; (Bottom Right): The first critical angular velocity vs. the temperatures from two different approaches.} }}\label{fig5}
\end{figure}
%******************

\emph{Single vortex solution in static}.--- We have investigated the dynamical evolution of the vortex lattices in the preceding sections. For completeness, in this section we will study the static single vortex solution in order to compare the first critical angular velocity $\Omega_{c1}$ to that obtained from dynamics. Analytically, $\Omega_{c1}$ can be obtained as \cite{Landau},
\begin{equation}
\Omega_{c1}=\frac{1}{M R^2} \ln\left(\frac{R}{a}\right),
\label{criticalomega}
\end{equation}
where $a$ denotes the vortex core size. { At the temperature $T=0.82T_c$ and $R=11$, we find that $a\sim 0.6257$ by fitting the single vortex order parameter as $\langle O(r)\rangle= \langle O(R)\rangle\tanh\left(r/(\sqrt{2}a)\right)$ \cite{Annett}, which is shown in the top left panel of Fig.\ref{fig5}. Substituting $M\sim0.8534$ from the above fitting Eq.\eqref{fitting} in to Eq.\eqref{criticalomega}, we get $\Omega_{c1}\sim0.0278$. By contrast, in the top right panel of Fig.\ref{fig5} we show the free energies of static $n=0$ and $n=1$ vortex solutions. The intersecting point $\Omega_{c1}\sim0.03$ is the first critical angular velocity, which is comparable to $\Omega_{c1}\sim0.0278$ (within $7.34\%$ differences) obtained from analytical derivation Eq.\eqref{criticalomega}. In the bottom panels of Fig.\ref{fig5} we also show the temperature dependence of the vortex size $a$ and the first critical angular velocity $\Omega_{c1}$. The time dependence of $a$ can also be vividly seen in the subfigures (D, E, F) in Fig.\ref{fig3}, in which as temperature grows the size of the each individual vortex grows as well. The fitted curve of this relation is $a(T)\sim0.2655\times(1-T/T_c)^{-1/2}$ as shown in the bottom left panel of Fig.\ref{fig5}. In the bottom right panel of Fig.\ref{fig5} we show the relation between $\Omega_{c1}$ and temperatures. The red curve and circles are from the intersections of the $n=0$ and $n=1$ free energies of single static vortex solutions as we already did above. The blue curve and triangles are from the formula Eq.\eqref{criticalomega} by substituting each $a(T)$ into this formula. We see that the critical angular velocities obtained from these two different approaches are consistent with each other in the sense that as temperature grows they will decrease, and their values are close to each other. }

\emph{Summary}.---
The vortex lattices were found in a holographic
superfluid in different temperatures for the first time.
We also confirm that the Feynman's linear relation is always satisfied in different temperatures. This was the advantage of holography than GP equation, which could not deal with finite temperature problem. The critical angular velocity from the analysis of the free energy also matched very well with theoretical predictions Eq.\eqref{criticalomega}.  We expect that this holographic study may shed light on understanding  the properties of a rotating strongly coupled superfluid.

%{\red Therefore, the vorticity equals to $\nabla \times \vec{v} =2\Omega$.
%Then the circulation can be computed according to Stokes' theorem
%$\Gamma=\oint \vec{v}\cdot d \vec{l} =2\Omega \pi R^2.$
%Since the circulation of a vortex is just the quantum
%number, thus the number of vortices can be easily determined from the condition that the
%velocity circulation along a contour enclosing a large number of vortices
%should  correspond to the rotation of the liquid as a whole. If
%such a contour encloses all the vortices along the circumference of  the disk, then the total number of vortices %$N$ admit the Feynman relation \cite{Feynman} as }

\emph{Acknowledgements}.---
We thank Christopher P. Herzog for valuable comments.
This work is supported by the National Natural
Science Foundation of China (under Grant No. 11675140, 11705005, 11875095, 11565017 and 11675015). Y.T. is also supported by the ``Strategic Priority Research Program of the Chinese Academy of Sciences'' with Grant No.XDB23030000.

H.B.Z. and H.Q.Z. share the equal contributions to this work as the corresponding authors.

%%%%%%%%%% Merge with supplemental materials %%%%%%%%%%
%\pagebreak
%\widetext
%\begin{center}
%\textbf{\large Supplemental Materials: Title for main text}
%\end{center}
%\section{Supplementary Material}


\vspace{0mm}
\begin{thebibliography}{99}
\vspace{0mm}
\bibitem{Abrikosov}  A. A. Abrikosov, Sov. Phys. JETP {bf 5}, 1174 (1957).
\bibitem{Tkachenko}   V. K. Tkachenko, Sov. Phys. JETP {\bf 22}, 1282 (1966); 23,
1049 (1966); 29, 945 (1969).


\bibitem{Packard} Richard E. Packard and T. M. Sanders, Jr., Phys. Rev. Lett., {\bf 22}, 823, (1969).
\bibitem{Yarmchuk} E.J. Yarmchuk, M.J.V. Gordon and R.E. Packard, Phys. Rev. Lett. {\bf 43} 214 (1979).
\bibitem{Bewley} G.P. Bewley, D.P. Lathrop and K. R. Sreenivasan,  Nature {\bf 441} 588 (2006).

\bibitem{Madison} K. W. Madison, F. Chevy, W. Wohlleben, and J. Dalibard,
Phys. Rev. Lett. {\bf 84}, 806 (2000).
\bibitem{Shaeer} J. R. Abo-Shaeer, C. Raman, J. M. Vogels, and W. Ketterle,
Science {\bf 292}, 476 (2001).

\bibitem{Pethick} C.J. Pethick and H. Smith, Bose-Einstein Condensation
in Dilute Gases, Cambridge University Press, Cambridge
(2002).
\bibitem{Fetter} A. L. Fetter,	Rev. Mod. Phys. {\bf 81}, 647 (2009).
\bibitem{Kasamatsu1} K. Kasamatsu, M. Tsubota, M. Ueda, Phys. Rev. Lett. {\bf 91} 150406 (2003).
\bibitem{Kasamatsu2} K. Kasamatsu, M. Tsubota, Phys. Rev. Lett. {\bf 97}, 240404 (2006).

\bibitem{Maldacena} J. M. Maldacena, Adv. Theor. Math. Phys. {\bf 2}, 231 (1998).
\bibitem{Gubser} S. S. Gubser, I. R. Klebanov, and A. M. Polyakov, Phys.
Lett. B {\bf 428}, 105 (1998).
\bibitem{Witten} E. Witten, Adv. Theor. Math. Phys {\bf 2}, 253 (1998).

\bibitem{Liu} A. Adams, P. M. Chesler, and H. Liu, Science {\bf 341}, 368
(2013).

\bibitem{Gubser2008} S. S. Gubser, Phys. Rev. D {\bf 78}, 065034 (2008).
\bibitem{Hartnoll} S. A. Hartnoll, C. P. Herzog, and G. T. Horowitz, Phys.
Rev. Lett. {\bf 101}, 031601 (2008).
%\cite{Herzog:2008he}
\bibitem{Herzog}
  C.~P.~Herzog, P.~K.~Kovtun and D.~T.~Son,
  %``Holographic model of superfluidity,''
  Phys.\ Rev.\ D {\bf 79}, 066002 (2009).
  %%CITATION = doi:10.1103/PhysRevD.79.066002;%%
  %226 citations counted in INSPIRE as of 01 Apr 2019

%\cite{Montull:2009fe}
\bibitem{Montull}
  M.~Montull, A.~Pomarol and P.~J.~Silva,
  %``The Holographic Superconductor Vortex,''
  Phys.\ Rev.\ Lett.\  {\bf 103}, 091601 (2009).
  %%CITATION = doi:10.1103/PhysRevLett.103.091601;%%
  %94 citations counted in INSPIRE as of 01 Apr 2019
%\cite{Dias:2013bwa}
\bibitem{Keranen}V. Keranen, E. Keski-Vakkuri, S. Nowling, and K. Yogendran, Phys.Rev. D 81, 126012 (2010).

\bibitem{Dias}
  Oscar~J.~C.~Dias, G.~T.~Horowitz, N.~Iqbal and J.~E.~Santos,
  %``Vortices in holographic superfluids and superconductors as conformal defects,''
  JHEP {\bf 1404}, 096 (2014).
  %%CITATION = doi:10.1007/JHEP04(2014)096;%%
  %39 citations counted in INSPIRE as of 01 Apr 2019
\bibitem{Wu}
  M.~S.~Wu, S.~Y.~Wu and H.~Q.~Zhang,
  JHEP {\bf 1605}, 011 (2016).

%\cite{Maeda:2009vf}
\bibitem{Maeda}
  K.~Maeda, M.~Natsuume and T.~Okamura,
  %``Vortex lattice for a holographic superconductor,''
  Phys.\ Rev.\ D {\bf 81}, 026002 (2010).
  %%CITATION = doi:10.1103/PhysRevD.81.026002;%%
  %91 citations counted in INSPIRE as of 01 Apr 2019

  \bibitem{Murata} K. Murata, S. Kinoshita, and N. Tanahashi, JHEP {\bf 1007},
050 (2010).
\bibitem{Bhaseen} M. J. Bhaseen, J. P. Gauntlett, B. D. Simons, J. Sonner,
and T. Wiseman, Phys. Rev. Lett. {\bf 110}, 015301 (2013).
%\cite{Li:2013fhw}
\bibitem{Li}
  W.~J.~Li, Y.~Tian and H.~b.~Zhang,
  %``Periodically Driven Holographic Superconductor,''
  JHEP {\bf 1307}, 030 (2013).
  %%CITATION = doi:10.1007/JHEP07(2013)030;%%
  %34 citations counted in INSPIRE as of 02 Apr 2019
%\cite{Bai:2014tla}
\bibitem{Bai}
  X.~Bai, B.~H.~Lee, L.~Li, J.~R.~Sun and H.~Q.~Zhang,
  %``Time Evolution of Entanglement Entropy in Quenched Holographic Superconductors,''
  JHEP {\bf 1504}, 066 (2015).
  %%CITATION = doi:10.1007/JHEP04(2015)066;%%
  %23 citations counted in INSPIRE as of 02 Apr 2019

\bibitem{Tianyu} X. Li, Y. Tian and H. Zhang, arXiv:1904.05497.

%\cite{Zeng:2016api}
\bibitem{Zeng:2016api}
  H.~B.~Zeng, Y.~Tian, Z.~Y.~Fan and C.~M.~Chen,
  %``Nonlinear Transport in a Two Dimensional Holographic Superconductor,''
  Phys.\ Rev.\ D {\bf 93}, 121901 (2016).
  %%CITATION = doi:10.1103/PhysRevD.93.121901;%%
  %14 citations counted in INSPIRE as of 02 Apr 2019
%\cite{Zeng:2016gqj}
\bibitem{Zeng:2016gqj}
  H.~B.~Zeng, Y.~Tian, Z.~Fan and C.~M.~Chen,
  %``Nonlinear Conductivity of a Holographic Superconductor Under Constant Electric Field,''
  Phys.\ Rev.\ D {\bf 95}, 046014 (2017).
  %%CITATION = doi:10.1103/PhysRevD.95.046014;%%
  %7 citations counted in INSPIRE as of 02 Apr 2019
%\cite{Zeng:2018ero}
\bibitem{Zeng:2018ero}
  H.~B.~Zeng and H.~Q.~Zhang,
  %``Universal critical exponents of nonequilibrium phase transitions from holography,''
  Phys.\ Rev.\ D {\bf 98}, no. 10, 106024 (2018).
  %%CITATION = doi:10.1103/PhysRevD.98.106024;%%
  %1 citations counted in INSPIRE as of 06 Apr 2019




%\cite{Du:2014lwa}
%\bibitem{Du:2014lwa}
%  Y.~Du, C.~Niu, Y.~Tian and H.~Zhang,
  %``Holographic thermal relaxation in superfluid turbulence,''
%  JHEP {\bf 1512}, 018 (2015).
  %%CITATION = doi:10.1007/JHEP12(2015)018;%%
  %17 citations counted in INSPIRE as of 02 Apr 2019
 %\cite{Chesler:2014gya}
% \bibitem{Chesler:2014gya}
%  P.~M.~Chesler, A.~M.~Garcia-Garcia and H.~Liu,
  %``Defect Formation beyond Kibble-Zurek Mechanism and Holography,''
%  Phys.\ Rev.\ X {\bf 5}, no. 2, 021015 (2015).
  %%CITATION = doi:10.1103/PhysRevX.5.021015;%%
  %32 citations counted in INSPIRE as of 02 Apr 2019
%\cite{Sonner:2014tca}
%\bibitem{Sonner:2014tca}
%  J.~Sonner, A.~del Campo and W.~H.~Zurek,
  %``Universal far-from-equilibrium Dynamics of a Holographic Superconductor,''
%  Nature Commun.\  {\bf 6}, 7406 (2015)
  %%CITATION = doi:10.1038/ncomms8406;%%
  %38 citations counted in INSPIRE as of 02 Apr 2019


\bibitem{Feynman} R. P. Feynman, Progress in Low Temperature Physics, {\bf 1},17-53, (1955).
\bibitem{Landau} L.D.Landau and E.M.Lifshitz, Statistical Physics, Part 2,(Oxford: Pergamon,1981).

%\cite{Natsuume:2017jmu}
\bibitem{Natsuume:2017jmu}
  M.~Natsuume and T.~Okamura,
  %``Kibble-Zurek scaling in holography,''
  Phys.\ Rev.\ D {\bf 95}, no. 10, 106009 (2017)
  %doi:10.1103/PhysRevD.95.106009
  %[arXiv:1703.00933 [hep-th]].
  %%CITATION = doi:10.1103/PhysRevD.95.106009;%%
  %3 citations counted in INSPIRE as of 11 Jun 2019
  %\cite{Domenech:2010nf}

\bibitem{Domenech}
  O.~Domenech, M.~Montull, A.~Pomarol, A.~Salvio and P.~J.~Silva,
  %``Emergent Gauge Fields in Holographic Superconductors,''
  JHEP {\bf 1008}, 033 (2010).
  %%CITATION = doi:10.1007/JHEP08(2010)033;%%
  %88 citations counted in INSPIRE as of 02 Apr 2019

%\bibitem{Campbell} L. J. Campbell, R. M. Ziff, Phys. Rev. B {\bf 20}, 1886 (1979).
%\bibitem{Castin} Y. Castin, R. Dum, Eur. Phys. J. D {\bf 7}, 399 (1999).



\bibitem{Ruutu} Ruutu, V. M., Parts, \"{U}., Koivuniemi, J. H., Kopnin, N. B. and
Krusius, M.  J. Low Temp. Phys. {\bf 107}, 93 (1997).

\bibitem{Peeters} M. V. Milosevic and F. M.
Peeters, Phys. Rev. Lett. 94, 227001
(2005).

\bibitem{Hess} G. B. Hess, Phys. Rev., {\bf 161}, 189 (1967).
%\bibitem{Pines} P. NoziÃÂ¡ÃÂ¡res and D. Pines, The Theory of Quantum Liquids
%(Addison-Wesley, Reading, MA, 1990).
%\bibitem{Lundh} E. Lundh, C. J. Pethick, H. Smith, Phys. Rev. A {\bf 55}, 2126
%(1997).


%\bibitem{Kasamatsu}K. Kasamatsu, M. Tsubota, M. Ueda, Phys. Rev. A {\bf 67}, 033610 (2003).
%\bibitem{Mingarelli} L. Mingarelli, R. Barnett,Phys. Rev. Lett. {\bf 122}, 045301 (2019).
\bibitem{Annett} J.F. Annett, ``Superconductivity, superfluids, and condensates", Oxford Master Series in
Condensed Matter Physics, first edition, Oxford University Press (2004).

%{\blue PNAS paper}

%{\blue Packard, first experiment on rotation}

\end{thebibliography}
\end{document}